\documentclass[12pt]{iopart}

\usepackage{epsfig} 
\begin{document}

\title[Archimedes Project]{The Archimedes project: a feasibility study for weighing the vacuum energy}

\author{E Calloni$^1$, S Caprara$^2$, M De Laurentis$^1$,G Esposito $^3$,M Grilli$^2$, E Majorana$^4$, G. P. Pepe$^5$, S Petrarca$^2$, P Puppo$^4$, F Ricci$^6$, L Rosa$^1$, C Rovelli$^7$, P Ruggi$^8$, N L Saini$^2$, C Stornaiolo$^3$,  F Tafuri$^9$}

\address{$^1$ Universit\`a di Napoli Federico II, 
Dipartimento di Fisica and 
INFN Sezione di
Napoli, Complesso Universitario di Monte S. Angelo,
Via Cintia Edificio 6, 80126 Napoli, Italy }
\address{$^2$ Universit\'a di Roma "La Sapienza"  P.le A. Moro 2, I-00185, Roma, Italy}
\address{$^3$ INFN  Sezione di
Napoli, Complesso Universitario di Monte S. Angelo,
Via Cintia Edificio 6, 80126 Napoli, Italy}
\address{$^4$ INFN  Sezione di
Roma, P.le A. Moro 2, I-00185, Roma, Italy}
\address{$^5$ Universit\`a di Napoli Federico II, 
Dipartimento di Fisica, piazzale Tecchio 80, 80126 Napoli, Italy}
\address{$^6$ Universit\'a di Roma "La Sapienza"  and INFN sezione di Roma, P.le A. Moro 2, I-00185, Roma, Italy}
\address{$^7$ Aix Marseille Universit\'e CNRS, CPT, UMR 7332, 13288
 Marseille, France \\
Universit\`e de Toulon, CNRS, CPT, UMR 7332, 83957 La Garde, France}
\address{$^8$ European Gravitational Observatory (EGO), 
I-56021 Cascina (Pi), Italy}
\address{$^9$ Dipartimento Ingegneria dell'Informazione, 
Seconda Universit\`a di Napoli, I-81031 Aversa (CE), Italy}

\ead{enrico.calloni@na.infn.it}
\vspace{10pt}
\begin{indented}
\item[]August 2014
\end{indented}

\begin{abstract}
 Archimedes is a feasibility study to a future experiment to ascertain the interaction of vacuum fluctuations with gravity. The future experiment should measure the force that the earth's gravitational field exerts on a Casimir cavity by using a balance as the small force detector. The Archimedes experiment analizes the important parameters  in view of the final measurement and experimentally explores solutions to the most critical problems.    
    
\end{abstract}

\pacs{04.80.Cc, 07.05.Fb}
%
\vspace{2pc}
\noindent{\it Keywords}: Vacuum Energy, Gravitation, High\_Tc superconductors, small force detectors
%
%
%
%

\section*{Introduction}

One of the profound open question of present physics is the irreconcilability among the quantum mechanical theory of vacuum and the General Relativity. The enormous value of the energy density of vacuum fluctuations as foreseen by quantum mechanics, if inserted in General Relativity theory is not at all compatible with the observed radius of the universe, nor with the acceleration of expansion: a problem known as the cosmological constant problem \cite{weinberg_art,Ishak}. At present, in spite of a detailed and important theoretical work, there is no general consensus on the theoretical solutions proposed \cite{Rovelli,Esposito,Kiefer} and on the fact that vacuum fluctuations do contribute to gravity \cite{rovelli2,why_not?}. Further, even if the common belief is that this should be the case no experiment has been done to finally verify or discard this assumption.  \\
In a recent paper we have shown that considering the present technological achievements on small force detectors, on superconductors and on seismic isolation it is possible to foresee an experimental path towards such a measurement \cite{archprd}. \\
The principle of the measurement is the weighing of a Casimir cavity. Indeed it can be shown that if a Casimir cavity is placed in the earth gravitational field and the vacuum energy does interact with gravity it receives a force directed upward equal to \cite{archprd,vacuum_fluctuation_force}:  
\begin{equation}   
\vec{F} =  {\cal A}\frac{\pi^2 \hbar }{720 a^3}\frac{\rm g }{c}\,\;\hat{z}
=\frac{E_{{\rm cas}}}{c^2}\,\vec{{\rm g}}.
\label{equazione_3}
\end{equation}

\noindent where ${\cal A}$ is the Casimir cavity proper area, $a$ is the Cavity proper distance among the plates, $c$ is the speed of light, $\vec{{\rm g}}$ is the earth gravitational acceleration (g its modulus), the unit vector $\hat{z}$ is directed upwards, $E_{{\rm cas}}$ is the Casimir energy and the evaluation is performed to first order with respect to the quantity $\frac{{\rm g \, a }}{c^2}$.
This force, directed upward, can be interpreted as the lack of weight of the modes that have been removed by the cavity, in similarity with the Archimedes buoyancy of fluid. Notice that, as expected (being assumed in the calculation  that the vacuum energy gravitates), the result is in agreement with the equivalence principle and the force can also be interpreted as the effect of the gravitational field on the negative mass associated to the Casimir energy.\\
In the light of present technologies the experimental verification could be approached with two main small force technologies: the gravitational wave detectors and the balances. The measurement principles and the experimental sensitivities of the two methods have been presented in \cite{archprd}. In the present paper we justify our choice of using balances and focus on the main experimental problems and the first solutions foreseen to reach the needed sensitivity. The paper is organized as follows: in section 1 the experimental scheme is presented, briefly recalling the use of a layered superconductor as Casimir multi-cavity to obtain a modulation of the signal. The role of the entropy in the measurement is also discussed. In section 2, the problem of seismic noise attenuation is discussed, and a particular strategy is proposed. In section 3 the thermal noise contribution is evaluated with respect to the critical parameters, the eventual thermal spurious modulation is evaluated and a solution proposed, and a dimensioning of the balance proposed.
Finally, in section 4, the attainable sensitivity is discussed and the main steps of the Archimedes project are presented.

\section{Layered superconductors as Casimir multi-cavities. The role of classical entropy}

The smallness of the force to be measured makes it mandatory to exploit the measurement with a modulation of the effect that brings the signal at frequencies within detectors measurement band. 
In \cite{kempf,archprd} it has been shown that layered superconductors, particularly the cuprates, are natural Casimir cavities, being structured as superconducting planes separated by dielectric planes.
Thus, the transition of a layered superconductor can be used to obtain a two-state modulation of the Casimir system that switches from a high (absolute value of) Casimir energy content in the superconducting phase to a low Casimir energy content when the superconductor is in the normal state.\\
An estimation of the variation of Casimir energy in the two states has been carried out in \cite{kempf,archprd} assuming that in the superconducting state the Casimir energy can be calculated within the zero-temperature and plasma infinitely thin sheets approximation, while it can be neglected in normal state due to the poor conductivity of the material in this state. In this approximation it can be shown that the Casimir energy $E_c(a)$ of two thin plasma sheets separated by the distance $a$ is equal to \cite{barton,bordag}
\begin{equation}
E_c(a) = -5 \times 10^{-3} \hbar \frac{c A}{ a^{5/2}} \sqrt{\Omega} .
\label{ekempf} 
\end{equation}
The parameter $\Omega$ is proportional to the density of the carrier in the plasma sheet \cite{barton,bordag}:
\begin{equation}
\Omega \equiv \frac{nq^2}{2mc^2\epsilon_0} ,
\end{equation}
where $n$ is the surface density of delocalized particles, $q$ their electric charge, $m$ their mass. 
In case of layered superconductors, particularly High-$T_{c}$ cuprates, the  particles' density is about $n = 10^{14} \, cm^{-2}$, the charge $q = 2e$, the mass $m = 2\alpha m_e$ with $\alpha = 5$.   Inserting these values in  Eq. (\ref{ekempf}), neglecting the Casimir energy in the normal state, considering a layered superconductor with typical distance $a$ = 1 nm and total thickness H, the variation of Casimir energy for unit volume is
\begin{equation} 
\Delta U_{{\rm cas}} \approx \eta(a) \frac{N \pi^2}{720}\frac{\hbar c}{a^{3}} \approx 2 \times 10^5 \,\, J/m^3 ,
\end{equation}
where $N \approx 10^9$ is the number of cavities per unit height. 
Remarkably, this variation is of the same order of magnitude of the total energy variation at the transition: Kempf hypothesis \cite{kempf} is here made, according to which the whole transition energy is actually Casimir energy. Nonetheless it is important to remark that, by virtue of the accuracy of the measurement, {\it even if the contribution of the Casimir energy were only of the order of the percent, its contribution to weight variation could be ascertained}.\\
Considering a volume of superconductor of the order of ten ${\rm cm}^3$, the corresponding Archimedes force is a weight variation of about $F_A \approx 10^{-16} \, N$. The force is tiny, but affordable from the most sensitive macroscopic detectors of small forces, like balances or Gravitational Wave (GW) detectors.  In a recent paper \cite{archprd} it has been shown that both of these systems could be suitable for detecting the Archimedes force from a sensitivity point of view. In particular, the third generation GW detectors, i.e. the planned Einstein Telescope, could reach a sensitivity of about $\tilde{F} = 3*10^{-15} N/\sqrt(Hz)$ in the low frequency region, corresponding  to the detection of the Archimedes force in tens of minutes of integration time \cite{et}.
In order to choose the detection system  one  experimental key point is the modulation of the effect, i.e. the periodical transition from normal to superconducting state. In particular, it must be compatible with the bandwidth of the detectors. The possible modulations of the transition are by temperature or external field.  Both are favoured in the low frequency region. This motivation leads us to the choice of the balances as the system to be experimentally used for the detection of the force.
Indeed in case of balances the detection bandwidth can be in the region of 1-100 mHz, comparable with torsion pendulums, while in case of Gravitational Wave detectors it extends from 10 Hz to few KHz. The possibility to modulate at so small frequency  is seen as a decisive argument in favour of using balances, even if in the long term the use of third generation GW detectors could be re-considered.\\
In exploiting the first experimental tests, particularly interesting is the role of the classical entropy in the weight measurement. In the following  classical entropy is meant to be the entropy as calculated for an ideal superconductor, disregarding completely the contribution of Casimir effect. This is to show that even in case that, contrary to previous evaluations and expectations \cite{kempf,archprd}, the Casimir effect were completely negligible, the proposed experiment would perform the interesting measurement of the weight of the entropy times the temperature.    
To demonstrate this let us consider the transition of type II superconductors of critical field ${\rm B_0}$ and critical temperature ${\rm T_c}$  obtained at fixed temperature $T$ by applying an external magnetic field. The transition is of the second order, with no latent heat. The weight variation will be the variation of internal energy $\frac{\Delta U}{c^2}g$, where U is the internal energy.  The variation of internal energy $\Delta U$ is given by
\begin{equation}
\Delta U   =  G_n(T) + TS_n(T) - G(T,0) - TS_s(T,0).  
\end{equation} 
The difference in Gibbs free energy is by definition
\begin{equation}
G_n(T) - G_s(T,0) = \frac{1}{2\mu_0}B_c(T)^2
\end{equation}
where  $B_c(T)$ is the critical field at the temperature T.
If  the relation   $B_c(T) = B_0\left[ 1 - \left( \frac{T}{T_c}\right )^2 \right]$ is assumed (valid for BCS superconductors but only approximately for layered type II superconductors), the entropy difference among normal and superconducting state at a given temperature is given by 
\begin{equation}
S_n(T) - S_s(T) = 2 \frac{B_0^2}{\mu_0}\left(\frac{T}{T_c^2}\right) \left[1 - (T/T_c)^2\right]
\end{equation}

and the internal energy variation can be expressed as

\begin{equation}
\Delta U =  \frac{B_0^2}{2\mu_0}\left[1 - (T/T_c)^2\right]^2 
+ 2 \left(\frac{T}{T_c}\right)^2 \left[1 - (T/T_c)^2\right].
\label{intEnergy} 
\end{equation}

\begin{figure}
\includegraphics[width=0.8\linewidth]{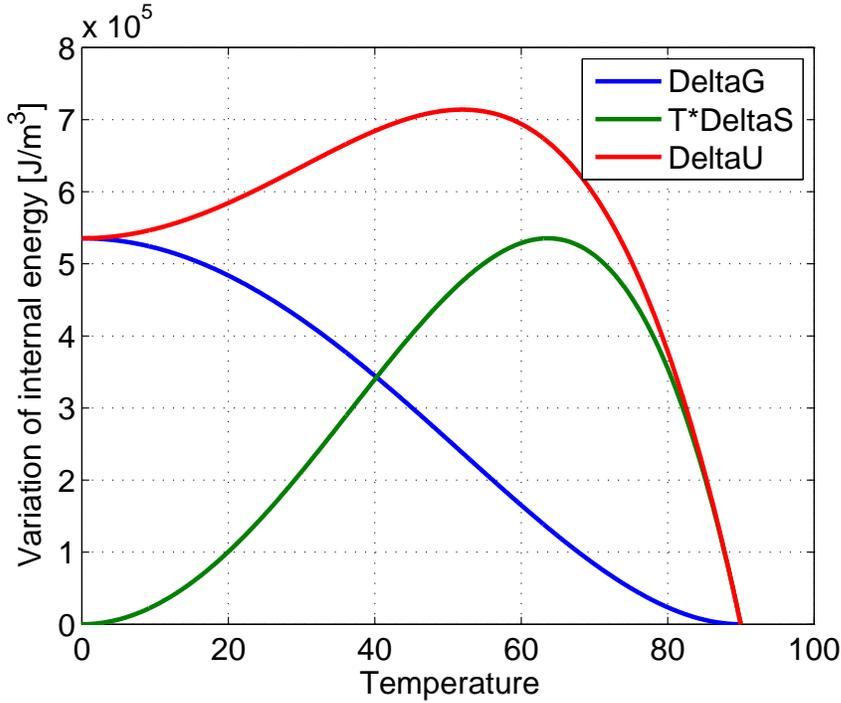}
\caption{Internal energy variation contributions in the approximation of parabolic critical field dependence from temperature T for a superconductor having $T_c$ = 90 K and thermodinamical critical zero field B0 = 1.16 Tesla  (YBCO typical values): the contribution due to entropy variation (DeltaS), the Gibbs free energy (DeltaG), the total (DeltaU)}
\label{entropy}
\end{figure}

Interestingly, if the transition is performed at a temperature not too far from $T_c$, the contribution of the entropy variation (multiplied by T) to internal energy variation is larger than the contribution to Gibbs free energy, as shown in figure \ref{entropy}. To our knowledge, whenever classically there is no doubt that the entropy does contribute to weight and mass through the temperature, no direct measurement of this contribution has been performed as yet. Thus, disregarding in this particular discussion the contribution of Casimir effect, this can be considered as a interesting side-measurement of the final experiment.  
    
\section{Seismic noise reduction}

\begin{figure}
\includegraphics[width=0.8\linewidth]{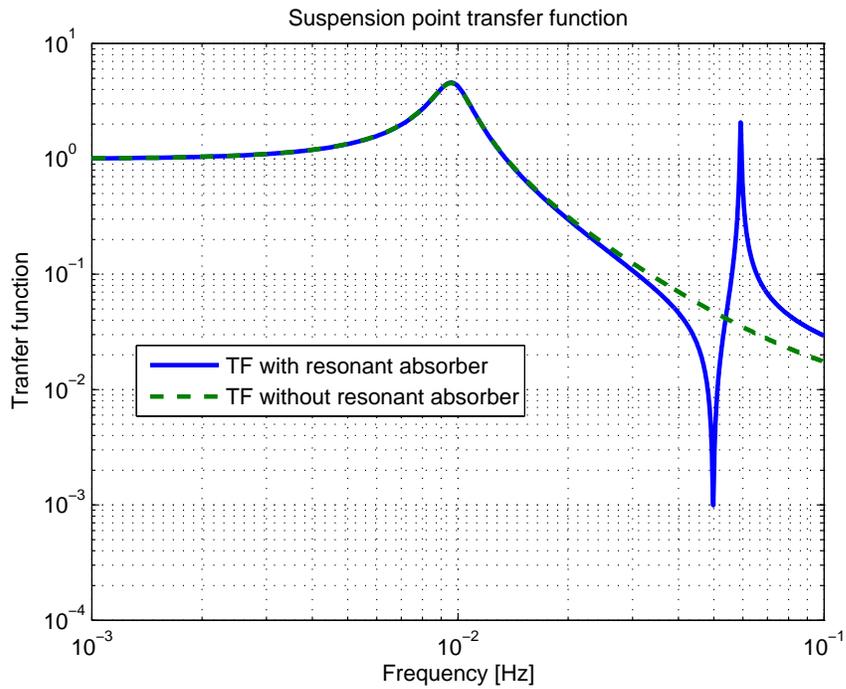}
\caption{Transfer function of seismic noise to the balance suspension point. The resonant absorber, continuous line, shows the complex zero-poles behavior that, at the complex zero frequency, reduces further the seismic noise for the zero-anti resonance quality factor }
\label{SeismicNoiseTF}
\end{figure}

\begin{figure}
{\center \hspace{2in} \includegraphics[width=0.2\linewidth]{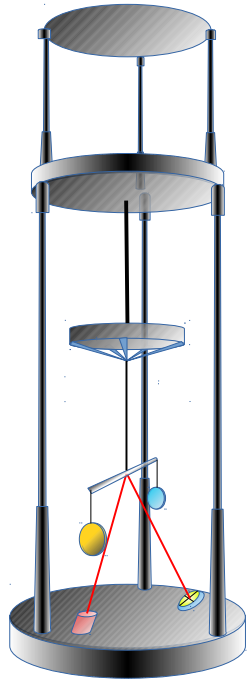}}
\caption{Schematic picture of the Archimedes force measurement. On the top of the inverted pendulum there is a second inverted pendulum acting as an absorbing stage. The balance is hanged to the intermediate spring-attenuation element. The signal is read by an optical lever system}
\label{2invPend}
\end{figure}

One of the main problems to be addressed in realizing a balance capable of measuring forces of the order of $10^{-16}$ N is the lack of an attenuation system in the very low frequency regime of 1-100 mHz. One possible strategy, already indicated in \cite{archprd}, is to hang the balance to a seismic isolation cascade formed by an inverted pendulum and blade-spring attenuator similar to the ones used in the Virgo gravitational wave detector. The inverted pendulum provides the attenuation in the two orizonthal translational degrees of freedom, while the blade-spring element takes care of the remaining degrees of freedom, vertical and rotationals.
The inverted pendulum has demonstrated a resonance frequency of 30 mHZ and studies are on going to further lower it to 10 mHz. Similar region of resonance frequency has been demonstrated for the blade-spring element.
Whenever these elements are at the best of present technology, they are still not sufficient to assure a sufficient attenuation in the mHz region needed for the Archimedes experiment. A possible solution could be the use of accelerometric sensors, placed on the top of the inverted pendulum, to be used in feed-back with unity gain above the Hz. This will reduce the inverted pendulum motion at the  electronic noise floor of the accelerometers 
$a_s \approx 4 \times 10^{-10} m^{2}/s \sqrt(Hz)$,  corresponding to the displacement  noise of  $1nm/\sqrt(hz)$ at 0.1 Hz, and flat for frequency 
less than 0.1 Hz \cite{harms}: if reached, this limit would be sufficient for the Archimedes force detection \cite{archprd}. The realization of such a feed-back system is quite complex and expensive, so that alternative passive solutions can be pursued. Here we present a passive solution based on a mechanical resonator, placed on the top of the Inverted-Pendulum and coupled to the Inverted-Pendulum so as to have the usual pair complex-zero/complex pole  tuned so that the complex-zero frequency is the same as the signal modulation frequency.
In this way the seismic energy is absorbed by the resonator at that frequency  and the suspension motion at the frequency is reduced by the attenuation transfer function of the Inverted-Pendulum and by the quality factor of absorber resonance. This behavior can be appreciated by looking at the transfer function of the seismic noise to the balance suspension point in figure \ref{SeismicNoiseTF}, obtained for the complete system as illustrated in figure \ref{2invPend}.

The seismic noise can vary in a remarkable way from site to site. In particular, in the recently tested very-low frequency environment of the Sos-Enattos mine, in Sardinia \cite{naticchioni}, in the region of frequency from 20 to  50 mHz it reaches a broad minimum of about $\tilde{a}_n \approx  10^{-8} \, \frac{m}{s^2}\frac{1}{sqrt{Hz}}$.

The coupling of suspension point acceleration $a_s$ can be interpreted as producing a moment of inertia $M_s = M_b\cdotp a_s \cdotp h_b$, where $M_b$ is the balance mass, $h_b$ is the balance bending point, equal to the distance among the center of mass and the center of rotation of the balance. This moment of inertia is  equivalent to the noise force  $ \tilde{F_n} = M_b \cdotp a_s \cdotp h_b/L_b$. The bending point determines the balance's resonance frequency $\omega_b$, with the relation $\omega_b^2 = \frac{M_b g\,h_b}{I}$. This distance can be tuned both mechanically, by regulating ballasts' position, and in feed-back, with the help of external forces. 
To calculate the expected signal and noises at the balance, we have 
considered a balance having arms of length L = 0.1 m, a plate at each 
arm's end of mass $M = 0.4$ kg, total mass $M_b = 1.25$ kg, moment of inertia 
$I = 0.01 \rm{kg \, m^{2}}$. The resonance frequency is placed in the region of low seismic noise: $F_{\rm{res}} = \omega_b/2\pi = 40$ mHz, with mechanical internal loss angle $\phi = 10^{-6}$.
Setting the resonance frequency in the tens of mHz frequency region makes it possible to relax the constraint on the accuracy of setting the bending point: in particular, the resonance of 40 mHz corresponds 
to the setting of the bending point distance from the balance center of 
mass of about $h_b = 50 \mu m$.  Within this design and seism conditions the equivalent noise force due to seismic noise is

\begin{equation}
{\tilde F_s} =  M_b \cdotp a_s \cdotp TF \cdotp h_b/L_b = 6\cdotp 10^{-15} \, N/\sqrt(Hz)  
\end{equation}

This value is compatible with the expected thermal noise \cite{archprd} and might allow a remarkable simplification of the system, not requiring the active system of accelerometers and control loops. Furthermore, the setting of the bending point in the tens of microns region is an easy task. \\ On the other hand, this choice requires a very quiet seismic environment, a condition which is not necessary if the Inverted Pendulum is controlled by an active loop.
The choice among the two solutions is part of the programme of the Archimedes feasibility study.

\section{Temperature modulation}

Temperature modulation in zero external magnetic field is the other no latent-heat transition from superconducting to normal state that can be used to modulate the effect. In can be shown \cite{archprd} that the variation of internal energy is the same of \ref{intEnergy} with the addition of the normal state contribution:

\begin{equation}
\Delta U = \int_T^{T_c}C_ndT + \frac{B_0^2}{2\mu_0}\left[1 - (T/T_c)^2\right]^2 
+ 2 \left(\frac{T}{T_c}\right)^2 \left[1 - (T/T_c)^2\right].
\label{dufinal}  
\end{equation}

Once more, the variation of internal energy is proportional to, and roughly of the same order of magnitude as, the energy of the thermodynamical critical field, but in this case the contribution of specific heat in normal state must be taken into account.
For superconductors whose transition temperature is in the tens-Kelvin region, this contribution can be neglected in the measurement. For superconductors in the one hundred K region of transition temperatures the phonon contribution becomes the major contribution to internal energy variation. One possible way of taking it into account is to subract it off-line, during the analisys of data.
A more efficient method is to subract it, at least at the leading order, directly with the measurement. This can be done equipping the balance with  equal superconductors on both size, as in figure \ref{2invPend}. Hence, the modulation on the first superconductor is $T_1(t) = T0 - \Delta T + A sin(\omega_m t)$ while on the second superconductor is  $T_2(t) = T0 + \Delta T + A sin(\omega_m t)$. The amplitude A is set lower than $\Delta T$ so that the first sample explores the superconducting transition while the second remains in normal state, and the balance measures the difference of weight due to superconducting contribution. The phononic part, which is not equally strongly dependent on temperature, is suppressed,  the modulation of weight being very similar in the two arms.  This can be appreciated by considering equation \ref{dufinal} developed for $\Delta T << T0$.  The amplitude of the signal $S_s = \Delta U_1 - \Delta U_2 $ at the modulation frequency due to the superconductor is 
\begin{equation}
S_s = \frac{4B_0^2}{\mu_0}\frac{dT}{T0}.
\end{equation}
The contribution of the phononic part is given at the first order and for the same frequency by
\begin{equation}
S_ph = 2\frac{\partial C_n(T)}{\partial T}\Delta T\cdot dT.
\end{equation}
The amplitude of the temperature modulation $dT$ will depend on the superconductor considered, and superconductors with lower specific heat are favoured. In case of YBCO, for example, the temperature modulation can be of few degrees, while the phononic contribution remains approximately negligible. This results in a modulation of the weight of about one half of the maximum reachable, as can be seen by figure \ref{entropy}, a condition that is acceptable from the experimental point of view. It is part of the Archimedes project to select the superconductors with best parameters with respect to the temperature modulation signal.\\
From a structural point of view the temperature modulation requires that the mass of the superconductor will be suspended to the arm of the balance, as in figure \ref{2invPend}. In this way the application point of the force does not change even in presence of small modification of the superconductor volume (due to thermal expansion/contraction), and spurious temperature modulation of the arm can be minimized.

\begin{figure}
\includegraphics[width=0.8\linewidth]{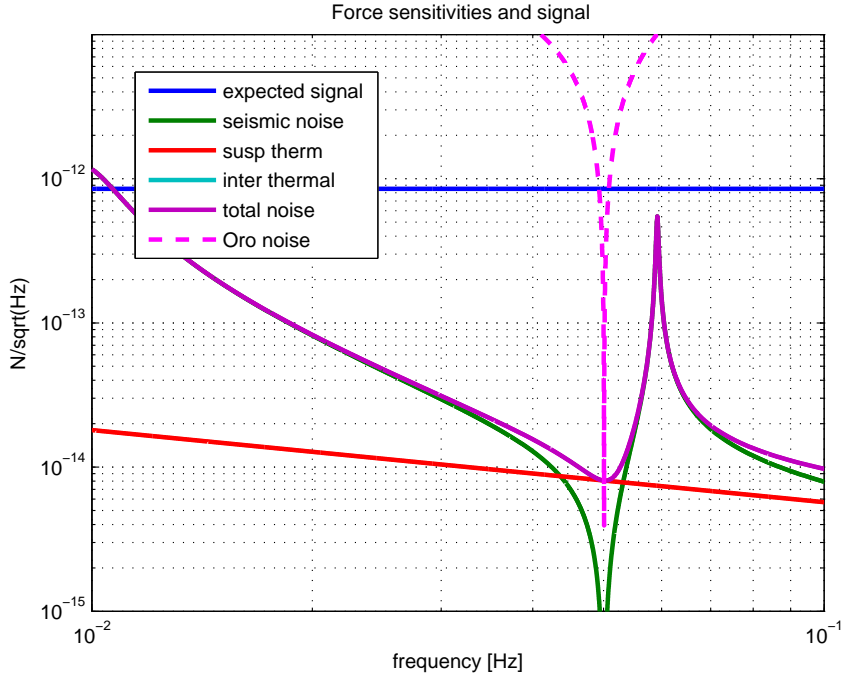}
\caption{Expected signal and noises. The detection bandwidth is within the resonance of the balance, limited by the read-out noise indicated as oroNoise}
\label{sensitivityForce}
\end{figure}

The major sources of noise in this scheme will be represented by the thermal noise and the seismic noise, as reported in figure \ref{sensitivityForce}. Considering a 250 $\mu m$ thick superconductor deposited on both faces of a disk of 0.15 m radius under the Kempf hypothesis that whole transition energy is due to Casimir effect, the Archimedes force would be $F_A = 4\cdot 10^{-16} \, N$.  
The read-out system, as discussed in \cite{archprd} and also reported in figure \ref{2invPend}, can be an optical lever.
As shown in figure \ref{sensitivityForce} with these choices the modulation frequency, the seismic-resonant absorber frequency and the balance frequency must be carefully  tuned to be the same and forced to be in the low-noise seismic region. Further, the experiment must be performed in very low-seismic noise sites, like the Sos-Enattos mine. These limitations are to be compared with the use of an active seismic noise reduction scheme, that in principle does not requires a so quiet environment, relaxes the constraints on the bandwidth, but at the price of a remarkable higher complexity \cite{archprd}. The assessment of the passive reduction and comparison of these two methods will be one of the major tasks of the Archimedes experiment.

\end{document}